\documentclass[twocolumn]{aastex631}

\usepackage{multirow}
\usepackage{amsmath}
\usepackage{physics}
\usepackage{pifont}

\begin{document}

\title{The In Situ Growth of Stellar-mass ``Light'' Seed Black Holes in Nuclear Star Clusters}

\correspondingauthor{Yanlong Shi}
\email{yanlong@cita.utoronto.ca}

\newcommand{\cita}{Canadian Institute for Theoretical Astrophysics, University of Toronto, Toronto, ON M5S 3H8, Canada}

\author[0000-0002-0087-3237]{Yanlong Shi}
\affiliation{\cita}

\author[0000-0002-8659-3729]{Norman Murray}
\affiliation{\cita}

\newcommand{\revision}[1]{\textcolor{red}{#1}}




\begin{abstract}

Remnant black holes (BHs) of massive stars (``light seeds'') are a potential origin for supermassive black holes (SMBHs). We use magnetohydrodynamic simulations to study the formation and growth of light seeds in star-forming giant molecular clouds (GMCs) with masses $10^5$--$10^9\,M_\odot$, which evolve for $\sim 10$--$30\,\rm Myr$ and form compact star clusters, akin to high-redshift nuclear star clusters. In particular, the simulations resolve very massive stars (VMSs, 100--$300\,M_\odot$), including their radiative and mechanical feedback, and model feedback-regulated accretion onto remnant BHs. We find that, even in compact GMCs capable of forming deep potential wells, the gas reservoir is expelled by sustained stellar feedback and rapidly dispersed after supernova explosions. Remnant BH populations emerge $\sim 3\,\rm Myr$ after the starburst and concentrate at the cluster center (where $\rho_{\rm BH}\sim 10^4$--$10^6\,M_\odot\,{\rm pc}^{-3}$). With our fiducial sub-grid BH accretion/feedback model, in-situ BH accretion is inefficient for forming heavy seeds: some direct-collapse BHs briefly accrete at $\sim$(1--10)$\times$ the Eddington rate, but they reach only $\sim 400$--$500\,M_\odot$. A top-heavy initial mass function or natal kicks do not change this conclusion. Runaway accretion is only possible if the sub-grid BH model allows a high fraction of Bondi inflow to reach the BH, in which case a few seeds can grow to $\sim 10^6\,M_\odot$. We also discuss multiple-generation star formation that may be intrinsically correlated with remnant BH accretion.

\end{abstract}



\section{Introduction} \label{sec:intro}

Recent observations have revealed the existence of supermassive black holes (SMBHs) powering quasars at high redshift \citep{WangYangFan_2021ApJ...907L...1W,KokorevCaputiGreene_2024ApJ...968...38K,TripodiMartisMarkov_2025NatCo..16.9830T}. However, the physical origin of these SMBHs and their progenitors, or ``seed BHs,'' remains poorly constrained \citep{InayoshiVisbalHaiman_2020ARA&A..58...27I,VolonteriHabouzitColpi_2021NatRP...3..732V}. Several scenarios have been proposed, including the direct collapse of pristine gas in atomic-cooling halos \citep{BrommLoeb_2003ApJ...596...34B,VolonteriRees_2005ApJ...633..624V,BegelmanVolonteriRees_2006MNRAS.370..289B,ChonHosokawaOmukai_2021MNRAS.502..700C}, runaway stellar mergers in dense star clusters \citep{PortegiesZwartBaumgardtHut_2004Natur.428..724P,GurkanFreitagRasio_2004ApJ...604..632G,ShiGrudicHopkins_2021MNRAS.505.2753S}, and the remnants of massive stars \citep{JohnsonBromm_2007MNRAS.374.1557J,TanakaHaiman_2009ApJ...696.1798T,ChonOmukai_2020MNRAS.494.2851C}. The last scenario requires that the seed black holes subsequently experience super-Eddington accretion to explain the highest redshift quasars.

The first two scenarios produce intermediate-mass black holes \citep[IMBHs;][]{GreeneStraderHo_2020ARA&A..58..257G}. Direct collapse yields ``heavy seeds'' of $\sim 10^3$--$10^5\,M_\odot$, whereas ``light seeds'', or  massive stellar remnants, have $M_{\rm BH}\sim 10^2\,M_\odot$. The discovery of $\sim 10^9\,M_\odot$ SMBHs by $z\sim 7$ favors heavy seeds if assuming Eddington-limited accretion \citep[i.e., $\dot M_{\rm BH} \lesssim M_{\rm BH}/t_{\rm Sal}$, where $t_{\rm Sal}\equiv 0.1 \kappa_{\rm es}c/(4\pi G) \sim 45\,\rm Myr$;][]{WangYangFan_2021ApJ...907L...1W}. However, the light-seed scenario remains possible if those seeds undergo super-Eddington growth. Both observations \citep[e.g.,][]{SuhScharwachterFarina_2025NatAs...9..271S} and simulations \citep{JiangStoneDavis_2014ApJ...796..106J,JiangStoneDavis_2019ApJ...880...67J} demonstrate that accretion can exceed the Eddington limit, while cosmological zoom-in simulations further show that such phases can occur in galaxies \citep{HopkinsGrudicSu_2024OJAp....7E..18H,HopkinsSquireSu_2024OJAp....7E..19H,HopkinsSuMurray_2025OJAp....8E..48H}. 

The growth of light seeds has been explored extensively, with mixed conclusions. Early formation at $z\gtrsim 20$ combined with sustained accretion could enable light seeds to reach SMBH masses \citep{TanakaHaiman_2009ApJ...696.1798T,MadauHaardtDotti_2014ApJ...784L..38M}. However, radiative and mechanical feedback from both progenitor stars and the BH itself may evacuate gas and suppress further growth \citep{JohnsonBromm_2007MNRAS.374.1557J,AlvarezWiseAbel_2009ApJ...701L.133A}. For example, simulations including supernova (SN) feedback find that only a small fraction of light seeds grow beyond $\sim 10^4\,M_\odot$ \citep{MehtaReganProle_2024OJAp....7E.107M}.

Recent work has shown that efficient accretion onto light seeds may occur in star-forming giant molecular clouds (GMCs) \citep{ShiKremerGrudic_2023MNRAS.518.3606S,ShiKremerHopkins_2024A&A...691A..24S}, particularly at high surface densities ($\Sigma \sim 10^4\,M_\odot\,\rm pc^{-2}$), in which case self-gravity confines dense gas even after cluster formation \citep{GrudicHopkinsFaucher-Giguere_2018MNRAS.475.3511G}. Moreover, the deepening potential well of a forming globular cluster can further enhance gas inflow toward centrally located BHs \citep{ShiKremerHopkins_2024ApJ...969L..31S}. In these scenarios, only $\sim 0.1$--$1\%$ of seed BHs experience rapid growth, but the large number of seeds formed in young massive clusters suggests that some may become massive \citep{ReganVolonteri_2024OJAp....7E..72R}.

These simulations assume \emph{pre-existing} seed BHs embedded in such environments without explaining their origin. This motivates the following question: can light seeds grow rapidly if they form \emph{in-situ} as remnants of massive stars? {Stellar feedback, especially that from SNe, can prevent rapid growth \citep[e.g.,][]{DuboisVolonteriSilk_2015MNRAS.452.1502D,HabouzitVolonteriDubois_2017MNRAS.468.3935H,MehtaReganProle_2024OJAp....7E.107M,ShinSmithSijacki_2026MNRAS.548ag580S}}, but dense gas can be trapped by massive and compact star clusters, as seen in previous simulations \citep{ShiKremerHopkins_2024ApJ...969L..31S,PartmannNaabLahen_2025MNRAS.537..956P}. Observationally, \citet{PascaleDaiMcKee_2023ApJ...957...77P} and \citet{PascaleDai_2024ApJ...976..166P} showed evidence of dense gas in the LyC cluster and the ``Godzilla'' cluster of the Sunburst Arc. Hence, the rate and duration of light-seed growth are uncertain. 

Addressing this question requires simulations to self-consistently model: (1) formation of stars, especially very massive stars (VMSs; $\gtrsim 100\,M_\odot$); (2) VMS evolution and feedback; (3) remnant BH formation; and (4) subsequent BH accretion. In this work, we match these requirements with a modified version of the FIRE-3 framework \citep[Feedback In Realistic Environments;][]{HopkinsKeresOnorbe_2014MNRAS.445..581H,HopkinsWetzelKeres_2018MNRAS.480..800H,HopkinsWetzelKeres_2018MNRAS.477.1578H,HopkinsWetzelWheeler_2023MNRAS.519.3154H}, as described in \S~\ref{sec:method}. We run simulations of star-forming cloud complexes, each evolved for $\gtrsim 10\,\rm Myr$, exceeding the lifetime of a $100\,M_\odot$ VMS. The simulation suites include variations in GMC mass, radius, and metallicity, as well as stellar initial mass function (IMF), natal kicks, and sub-grid BH feedback (\S~\ref{sec:results}). We discuss the implications of our findings in \S~\ref{sec:discussion} and summarize our conclusions in \S~\ref{sec:summary}.

\begin{figure}
    \centering
    \includegraphics[width=\linewidth]{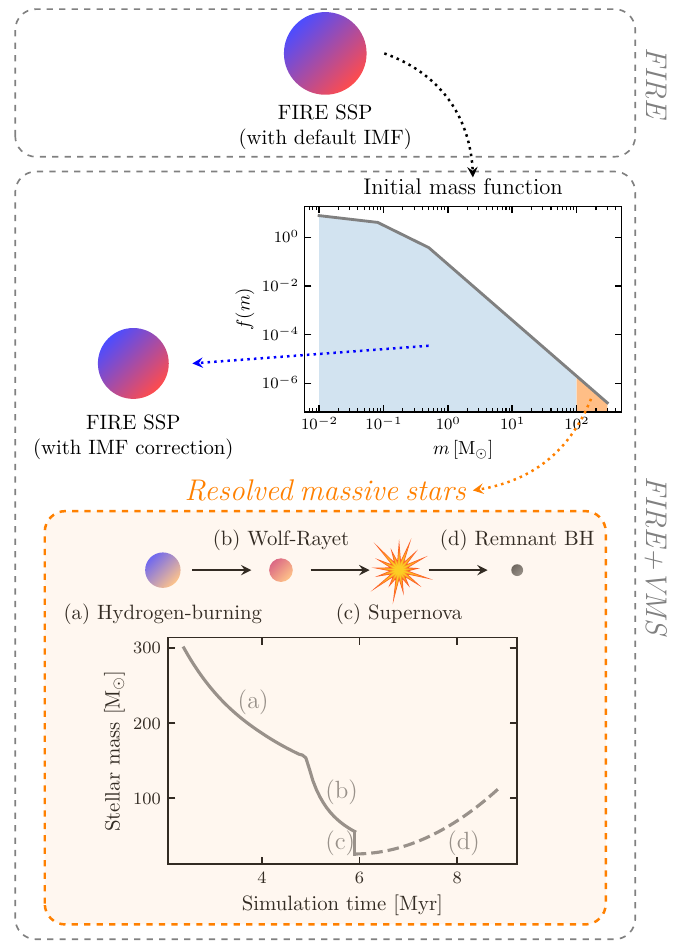}
    \caption{Treatment of stellar evolution and feedback in this work \citep[also see][]{ShiDaiMurray_2026ApJ...997..309S}, which splits resolved VMSs (with $m_{\rm ZAMS}>m_{\rm cut}$, here $m_{\rm cut}=100\,M_\odot$) from FIRE SSPs following the IMF. The VMS sub-grid model tracks four phases: (a) hydrogen-burning, (b) Wolf-Rayet, (c) supernova, (d) remnant BH formation and feedback-regulated accretion. Meanwhile, FIRE SSPs evolve with an IMF correction to avoid double-counting of stellar feedback.
    }
    \label{fig:diagram}
\end{figure}

\section{Method} \label{sec:method}

We run numerical simulations with {GIZMO}, a Lagrangian, meshless, Godunov hydrodynamical and $N$-body code \citep{Hopkins_2015MNRAS.450...53H}, in its meshless finite-mass (MFM) mode. The governing equations are the standard magnetohydrodynamical (MHD) equations as described in \citet{HopkinsRaives_2016MNRAS.455...51H}. On top of the MHD and gravity solver, we enable the FIRE-3 framework \citep{HopkinsKeresOnorbe_2014MNRAS.445..581H,HopkinsWetzelKeres_2018MNRAS.477.1578H,HopkinsWetzelKeres_2018MNRAS.480..800H,HopkinsWetzelWheeler_2023MNRAS.519.3154H} to model star formation and feedback by consolidating additional physics like cooling and multi-band radiation transfer \citep{HopkinsQuataertMurray_2012MNRAS.421.3488H,HopkinsWetzelKeres_2018MNRAS.480..800H,HopkinsGrudicWetzel_2020MNRAS.491.3702H}. The treatment has been applied to simulate star formation in GMCs \citep{GrudicHopkinsFaucher-Giguere_2018MNRAS.475.3511G}, resulting in properties of young massive clusters in fairly good agreement with observations \citep{GrudicGuszejnovHopkins_2018MNRAS.481..688G,GrudicHopkinsQuataert_2019MNRAS.483.5548G,GrudicHafenRodriguez_2023MNRAS.519.1366G}.

FIRE does not resolve individual stars that are essential to model the formation of remnant BHs: instead, each ``star particle'' represents a single stellar population (SSP), corresponding to an IMF-averaged ensemble of stars. While high-resolution star formation frameworks such as STARFORGE \citep{GrudicGuszejnovHopkins_2021MNRAS.506.2199G,GuszejnovGrudicHopkins_2021MNRAS.502.3646G} are capable of resolving individual stars in detail, they are currently computationally prohibitive for our purposes for different reasons. (1)~Star formation in GMCs proceeds on a characteristic free-fall timescale, $t_{\rm ff} = \left( 3\pi R_{\rm cl}^3 /32\,G\,M_{\rm cl} \right)^{1/2}$ \citep{ChevanceKrumholzMcLeod_2023ASPC..534....1C}, whereas the lifetimes of massive stars are $\gtrsim 3\,\rm Myr$; for the compact GMCs of interest here, $t_{\rm ff} \ll 3\,\rm Myr$. Evolving until the formation of remnant BHs would require many free-fall times. (2)~Compact GMCs can be extremely massive ($\gtrsim 10^{8}\,M_\odot$), making it impractical to simulate with STARFORGE (typical resolution is $\sim 0.01\,M_\odot$ per gas cell; \citealt{GrudicGuszejnovHopkins_2021MNRAS.506.2199G}).

Because our focus is on massive stars that produce remnant BHs within the first $\sim 10\,\rm Myr$, we adopt a ``hybrid'' approach that tracks the most massive ($M\gtrsim50$ or $100\,M_\odot$, see below) stars individually while treating all lower-mass stars as SSPs (Figure~\ref{fig:diagram}). The two stellar components are evolved using different prescriptions: massive stars are followed with dedicated sub-grid models, while SSPs evolve within FIRE (with appropriate corrections to account for the modified stellar content). This approach is conceptually similar to existing works \citep{LahenNaabJohansson_2020ApJ...891....2L,AnderssonAgertzRenaud_2023MNRAS.521.2196A,PartmannNaabLahen_2025MNRAS.537..956P,AnderssonReyPontzen_2025ApJ...978..129A}, though it is developed here to address a distinct physical problem. We describe our implementation below.

\subsection{FIRE+VMS}

We adopt a Kroupa-like piecewise IMF \citep{Kroupa_2002Sci...295...82K}, given by $\dd n/\dd m \propto m^{-0.3}$ for $0.01\,M_\odot \le m < 0.08\,M_\odot$, $\propto m^{-1.3}$ for $0.08\,M_\odot \le m < 0.5\,M_\odot$, and $\propto m^{-2.3}$ for $ 0.5\,M_\odot \le m \le 300\,M_\odot$. Because our simulations target at $\sim 10\,\rm Myr$, and only VMSs are short-lived enough \citep[$\gtrsim 3\,\rm Myr$; e.g.,][]{PortinariChiosiBressan_1998A&A...334..505P} to produce remnant BHs within this timescale, we explicitly resolve only the most massive stars. We adopt a mass threshold $m_{\rm cut}=100\,M_\odot$ (or $50\,M_\odot$ for GMCs with $M_{\rm cl}\le 10^6\,M_\odot$): stars with their zero-age main-sequence (ZAMS) masses above this threshold are extracted and treated as resolved VMS particles, otherwise they remain unresolved in FIRE SSPs.

For a FIRE SSP particle of mass $m_{\star,\rm FIRE}$, the average number of massive stars heavier than $m_{\rm cut}$ is $\mu = m_{\star,\rm FIRE}\, f_{\rm massive}/\langle m_{\rm massive} \rangle$, where $f_{\rm massive} = \int_{m_{\rm cut}}^{m_{\rm max}} m {dn/dm}(m)\,\dd m \big/ \int_{m_{\rm min}}^{m_{\rm max}} m {dn/dm}(m)\,\dd m$ is the mass fraction above $m_{\rm cut}$, and $\langle m_{\rm massive} \rangle$ is the mean stellar mass in this range. We draw the number of VMSs to split with Poisson sampling, and sample their ZAMS masses from the high-mass end of the IMF. The corresponding mass is subtracted from $m_{\star,\rm FIRE}$ to ensure mass conservation. In rare high-resolution cases where $m_{\rm ZAMS} > m_{\star,\rm FIRE}$, the entire SSP particle is converted into a single VMS. It is then merged with newly formed neighboring SSP particles (satisfying $|\mathbf x_j - \mathbf x_i| < \min(h_i, h_j)$ and forming within $0.1\,\rm Myr$, where $h_i$ is the smoothing length of particle $i$) until the target VMS mass is reached. Linear momentum and center-of-mass position are conserved during each merger.

When VMSs are treated separately, the remaining SSPs should no longer contribute feedback from the high-mass end of the IMF. So we renormalize the luminosity and mass-loss rates of SSPs using an updated light-to-mass ratio \citep[similar to][]{GuszejnovHopkinsMa_2017MNRAS.472.2107G}.


\subsection{VMS sub-grid model}
\label{sec:method:singe_star_feedback}

We develop the sub-grid model of VMSs based on public stellar evolution data PARSEC v2.0 \citep{NguyenCostaGirardi_2022A&A...665A.126N,CostaShepherdBressan_2025A&A...694A.193C}. The model was previously  used to study the chemical feedback from VMSs \citep{ShiDaiMurray_2026ApJ...997..309S}, which the reader may also consult.

\subsubsection{Mass evolution}
\label{sec:method:singe_star_feedback:mass}


PARSEC models provide the evolutionary track of each star with initial mass $m_{\rm ZAMS}$ and metallicity $Z$. We compile interpolation tables of useful quantities as a function of $(m_{\rm ZAMS}, Z)$, including: (1) the time of the {stellar wind mass loss rate} bistability jump, $t_{\rm kink}$; (2) the lifetime of the star, $\tau$; (3) the pre-SN mass, $m_{\rm preSN}$; (4) the remnant mass, $m_{\rm rem}$. 

The mass loss rate can be extracted from evolutionary tracks. For $t<t_{\rm kink}$, the best-fit mass-loss recipe is 
\begin{align}
    \log \dot m & = -13.3 + 6.4 \log(m) + 0.9 \log(Z/0.02) \nonumber \\ & -1.06\,\log(m)^2, \label{equ:wind_mass_loss_rate}
\end{align}
where $\dot m$ is the mass loss rate of main-sequence VMS in $M_\odot\,{\rm yr}^{-1}$, $m$ is the ``instantaneous'' stellar mass in $M_\odot$ (in contrast to the ZAMS mass $m_{\rm ZAMS}$). For $t_{\rm kink}<t<\tau$, we assume constant mass loss rate between $m(t_{\rm kink})$ and $m(\tau) = m_{\rm preSN}$. This treatment approximates, fairly accurately, the mass evolution over a wide range of $m_{\rm ZAMS}$ and $Z$ \citep[see Figure~4 of][]{ShiDaiMurray_2026ApJ...997..309S}.

Once the VMS star reaches its lifetime $\tau$, it becomes a ``supernova,'' then potentially a post-SN remnant \citep[depending on $m_{\rm ZAMS}$ and $Z$;][]{CostaShepherdBressan_2025A&A...694A.193C}. In particular, low-metallicity stars in the ``mass gap'' terminate as pair-instability supernovae (PISNe), leaving no remnants; above the PISN mass gap, remnants are direct-collapse BHs (DCBHs), and $m_{\rm preSN}$ is almost fully converted to $m_{\rm rem}$. We find DCBHs important for remnant BH accretion (\S~\ref{sec:results}).

\subsubsection{Feedback}

Stellar feedback in FIRE includes two types: radiative and mechanical \citep{HopkinsWetzelKeres_2018MNRAS.480..800H,HopkinsWetzelWheeler_2023MNRAS.519.3154H}. For our ``FIRE+VMS'' framework, we implement the VMS feedback by having these particles use standard FIRE feedback loops, but with mass-dependent details (e.g., luminosities, mass loss) are evaluated separately. Therefore, resolved VMSs have the same feedback effects simulated as they would contribute as part of FIRE SSPs.

The radiative feedback physics includes radiation transfer in three bands (IR, optical, UV), UV heating, HII heating, and radiation pressure \citep{HopkinsWetzelWheeler_2023MNRAS.519.3154H}. For VMSs, we modify the ``source term,'' i.e., the luminosities in different bands. The bolometric luminosity $L$ is fitted from the PARSEC data:
\begin{align}
    \log L = 0.0093 + 4.86 x - 1.13 x^2 + 0.114 x^3,
\end{align}
where $L$ is in $L_\odot$ and $x = \log(m_{\rm ZAMS}/{M_\odot})$. The luminosity in a radiative band is distributed following the blackbody spectrum \citep[][]{GrudicGuszejnovHopkins_2021MNRAS.506.2199G}, where the effective temperature is dependent on stellar radius, for which we use the fitting function in \citet{ToutPolsEggleton_1996MNRAS.281..257T}.

The VMSs' mechanical feedback (stellar and SN winds) is simulated as isotropic and momentum-driven winds following \citet{HopkinsWetzelKeres_2018MNRAS.477.1578H}, which conserves mass, momentum, and energy. The stellar wind is determined by the mass loss recipe described in \S~\ref{sec:method:singe_star_feedback:mass}, and the terminal velocity follows \citet{Vink_2018A&A...615A.119V}.
The SN wind carries $E_{\rm SN} = 10^{51}\,\rm erg$, so the terminal velocity is $ v_{\rm ej} = (2E_{\rm SN}/m_{\rm ej})^{1/2}$, where $m_{\rm ej} = m_{\rm preSN} - m_{\rm rem}$ is the ejecta mass. However, important exceptions are stars that terminate as DCBHs with $m_{\rm ej}\approx 0$,  for which we assume negligible SN feedback and so skip the SN feedback routine in the simulation.

\subsubsection{Dynamics}

Stellar-mass BHs may have natal kicks due to anisotropic mass loss or binary evolution \citep{ArcaSeddaMapelliBenacquista_2023MNRAS.520.5259A}; the distribution of kick velocities is uncertain. Using Gaia data, \citet{NagarajanEl-Badry_2025PASP..137c4203N} suggest that at least some BHs are born with strong natal kicks, up to $v_{\rm kick}\gtrsim 100\,\rm km\,{s}^{-1}$. Based on the dark remnant mass fraction of Milky Way clusters \citep{DicksonSmithHenault-Brunet_2024MNRAS.529..331D}, \citet{Rostami-ShiraziBaumgardtZonoozi_2025MNRAS.536.1332R} found that low natal kicks \citep[compared to $\sigma_{v} = 190\,\rm km\,{s}^{-1}$ for pulsars;][]{HansenPhinney_1997MNRAS.291..569H} are required. In this study, we assume that kick velocities are ``moderate,'' following a logarithm-uniform distribution in $[0.1,100]\,\rm km\,{s}^{-1}$.

VMSs are significantly more massive than field stars, causing non-negligible dynamical friction \citep{Chandrasekhar_1943ApJ....97..255C}. This friction makes massive stars ``sink'' to the center of the globular cluster \citep[e.g.][]{ShiGrudicHopkins_2021MNRAS.505.2753S}. In our simulations, since less-massive stars are collectively treated as FIRE SSPs, a direct $N$-body calculation misses the contribution of dynamical friction to VMS dynamics. To address this problem, apply an additional acceleration $\textbf{a}_{\rm df}$ acting on individual VMSs, so $\mathbf{a} = \mathbf{a}_{\rm grav} + \mathbf{a}_{\rm df}$, where $\mathbf{a}_{\rm df}$ is calculated with the sub-grid model as in \citet{MaHopkinsKelley_2023MNRAS.519.5543M}, which accounts for the contributions from all massive particles to the dynamical friction.

\subsection{Remnant BH accretion and feedback}

Unlike simulations in \citet{ShiKremerGrudic_2023MNRAS.518.3606S,ShiKremerHopkins_2024A&A...691A..24S}, the runs presented here typically do not fulfill $m_{\rm gas} \ll M_{\rm BH}$, making the gravitational capture algorithm used in those studies no longer applicable. Instead, we use the Bondi-Hoyle-Lyttleton (BHL) formula \citep{HoyleLyttleton_1939PCPS...35..405H,Bondi_1952MNRAS.112..195B} to evaluate BH accretion:
\begin{align}
    \dot M_{\rm Bondi} = 4\pi G^2 M_{\rm BH}^2 \rho/(c_{\rm s}^2+|\mathbf{v}_{\rm BH}-\mathbf{v}_{\rm gas}|^2)^{3/2}.
\end{align}

Following \citet{HopkinsWetzelKeres_2018MNRAS.477.1578H}, we implement a module of ``continuous'' BH accretion in {GIZMO}: instead of ``discretely'' swallowing an entire neighboring gas cell in a single timestep \citep{SpringelDiMatteoHernquist_2005MNRAS.361..776S,DiMatteoSpringelHernquist_2005Natur.433..604D} (which is not appropriate here since $M_{\rm BH} \lesssim m_{\rm gas}$), each BH (labeled $i$) receives a mass inflow of $\dot M_{\rm Bondi}\Delta t$ in a time bin $\Delta t$. Each of the interacting neighboring gas cells, defined with $|\mathbf x_j - \mathbf x_i |<{\rm min}(h_i, h_j)$, loses a mass of $-\Delta m_j = \dot M_{\rm Bondi}\Delta t \cdot \Delta \Omega_j /4\pi$ to maintain mass conservation, where $\Omega_j$ is the solid angle subtended by gas cell $j$ relative to the BH particle $i$. This, in principle, assumes isotropic inflow at the Bondi radius. The weight $\Delta \Omega_j /4\pi$ is evaluated the same way as that in \citet{HopkinsWetzelKeres_2018MNRAS.477.1578H}: $\Delta \Omega_j /4\pi\approx \tilde{w}_j \equiv w_j/\sum_k w_k $ and $w_j = [1-1/\sqrt{1+\mathbf{A}_{ji} \cdot \hat{\mathbf{x}}_{ji}/(\pi |\mathbf{x}_{ji}|^2) }]/2$, where $\mathbf{A}_{ji}$ is the effective contact surface area between particle $i$ and $j$, defined as $\mathbf{A}_{ji} = \hat{\mathbf{x}}_{ji} [\bar n_i^{-2} \partial W(r, h_i)/\partial r + \bar n_j^{-2} \partial W(r, h_j)/\partial r ]\vert_{r=|\mathbf{x}_{ji}|}$. We note that $w_j \sim \Delta \Omega_j / 4\pi $ \citep{HopkinsWetzelKeres_2018MNRAS.477.1578H}. The BH also gains momentum and angular momentum (AM) from neighbors, e.g., $\mathbf{p}_i \to \mathbf{p}_i + \sum_j|\Delta m_j|/m_j\cdot \mathbf{p}_j$. 

Only a fraction, $f_{\rm acc}$, of $\dot M_{\rm Bondi}$ should eventually reach the BH due to BH feedback. {The fraction $f_{\rm acc}$ is tiny ($\sim (r_{\rm g}/r_{\rm Bondi})^{1/2}$) for low-luminosity accretion flow \citep{GuoStoneKim_2023ApJ...946...26G,ChoPratherNarayan_2023ApJ...959L..22C} due to suppression from magnetic fields, }but could arguably be higher as suggested by simulations of slim-disk-like super-Eddington accretion flows \citep{JiangStoneDavis_2014ApJ...796..106J,ZhangStoneWhite_2026ApJ..1001..138Z}. In our fiducial simulations, we fix $f_{\rm acc}=0.05$; but we also include runs with variations in $f_{\rm acc}$ in \S~\ref{sec:discussion:feedback_test}.

The other portion of the inflow, $(1-f_{\rm acc})\dot M_{\rm Bondi} \Delta t$, is returned to the neighboring gas cells to account for BH mechanical feedback. The wind is bipolar and symmetric along the AM of gas within the BH's softening kernel, and $\dd \dot M_{\rm out}/\dd \theta \propto g (\theta) = s(s+\cos^2\theta)/[(s+1)(s+1-\cos^2\theta)]$ where $\theta$ is the polar angle relative to the AM and $s$ (=0.35) controls the width of the wind \citep{HopkinsTorreyFaucher-Giguere_2016MNRAS.458..816H}. Numerically, the mass outflow is coupled to gas neighbors with their weights ($w_j$) rescaled with $g(\theta)$. The terminal velocity of the feedback wind is a key factor but is not well constrained;  simulations \citep{ZhangStoneWhite_2026ApJ..1001..138Z} find a fast launching velocity of $\mathcal{O}(0.1)\,c$. Observations of X-ray binaries suggest wind velocities in the rough range of $\sim 1000$--$3000\,\rm km\,{s}^{-1}$ \citep{Munoz-DariasDiazTrigoDone_2026arXiv260105319M}, but some tentative measurements of high-ionization winds reach $\sim 10^4\,\rm km\,{s}^{-1}$ or even $\sim 4.5\times 10^4\,\rm km\,{s}^{-1}$ \citep{DelSantoPintoMarino_2023MNRAS.523L..15D}. We adopt a fiducial terminal velocity of $ 3000\,{\rm km\,{s}^{-1}} \approx 0.01\, c$, {but present results with other values (\S~\ref{sec:discussion:feedback_test}).}

BH radiative feedback is evaluated with a total luminosity {$\epsilon_{\rm r} \dot M_{\rm BH,0} c^2$ (where $M_{\rm BH,0}=f_{\rm acc} \dot M_{\rm Bondi}$ is the rest-mass inflow rate)} and distributed to different radiation bands following a template from \citet{ShenHopkinsFaucher-Giguere_2020MNRAS.495.3252S}, where $\epsilon_{\rm r}$ is the radiative efficiency. We employ the slim-disk model from \citep{MadauHaardtDotti_2014ApJ...784L..38M}, so $\epsilon_{\rm r} \sim 0.1$ at low accretion rates, but drops to $\sim 10^{-2}$ in the super-Eddington regime due to photon trapping. Specific radiative effects include photon momentum, HII heating, and Compton heating, following FIRE implementations introduced in \citet{HopkinsTorreyFaucher-Giguere_2016MNRAS.458..816H} and \citet{HopkinsWetzelKeres_2018MNRAS.480..800H}. 

{In the simulation, the actual BH mass growth rate ($\dot M_{\rm BH}$) is determined by the BHL accretion rate, as well as sub-grid feedback parameters corresponding to mass outflow ($(1-f_{\rm acc})\dot M_{\rm Bondi}$) and radiation loss in the rest mass ($\epsilon_{\rm r} \dot M_{\rm BH,0}$). We also allow super-Eddington accretion of $\le 1000\,\dot M_{\rm Edd}$, so finally $\dot M_{\rm BH} = \min [(1-\epsilon_{\rm r})f_{\rm acc}\dot M_{\rm Bondi}, 1000\,\dot M_{\rm Edd}]$.
}

\begin{figure*}
    \centering
    \includegraphics[width=\linewidth]{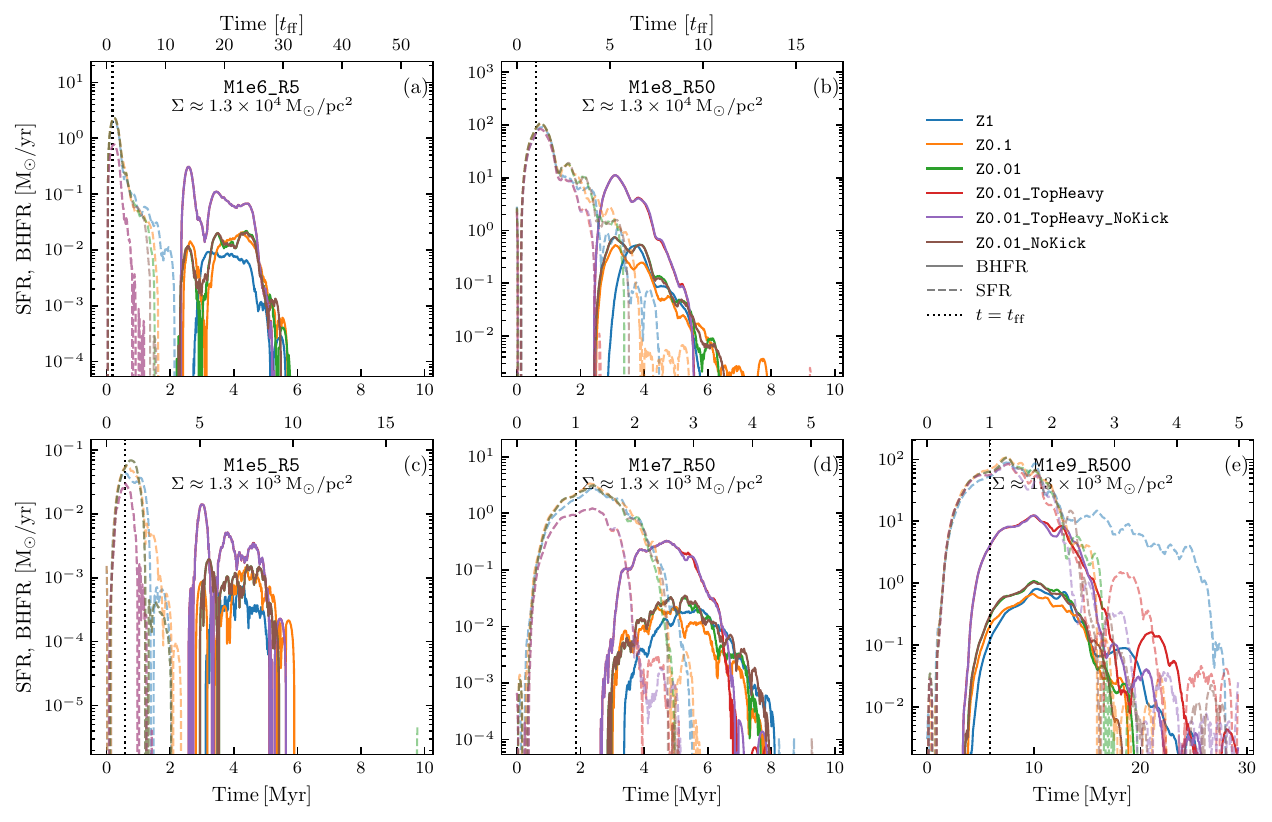}
    \caption{{Star formation rate (SFR; \emph{dashed}) and BH formation rate (BHFR; \emph{solid}) throughout the evolution of each simulation. Each panel represents a GMC complex listed in Table~\ref{tab:deluxesplit}, with the GMC's mass, radius, and surface density labeled. The colored lines represent the time evolution of SFR and BHFR of different simulation runs varying the GMC's initial metallicity ($0.01\,Z_\odot$--$Z_\odot$) and additional physics (IMF shape, natal kicks).
    We caution that BHFR is defined as the rate at which \emph{resolved} stars turn into remnant BHs at the end of their life in the simulation.} }
    \label{fig:sfr-bhfr}
\end{figure*}

\begin{figure*}
    \centering
    \includegraphics[width=\linewidth]{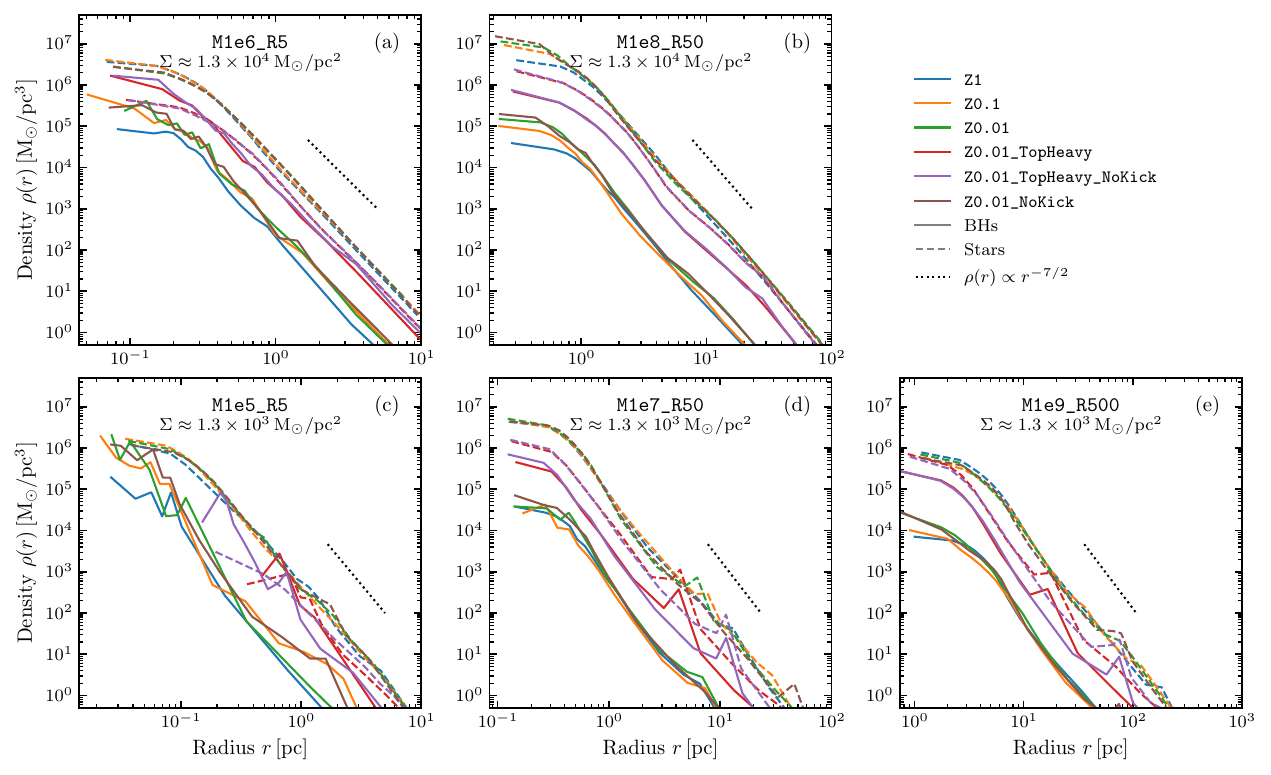}
    \caption{{Density profiles of stars (\emph{dashed}) and BHs (\emph{solid}) in the star cluster at the end of each simulation. Each panel represents a GMC complex in the same way as Fig.~\ref{fig:sfr-bhfr}, and the colored lines represent the more specified simulation runs varying metallicity, IMF shape, and natal kicks. Each density profile is calculated by setting the center $\mathbf{x}_{\rm center}$ from the minimum of the potential field, as a function of $r_{\rm cl} \equiv |\mathbf{x} - \mathbf{x}_{\rm center}|$.  Note that some curves are noisy since no regular-shaped globular clusters form in these simulations. We caution that ``BHs'' here are the remnants of \emph{resolved} stars in the simulation.}
    }
    \label{fig:density-profile}
\end{figure*}

\begin{figure*}
    \centering
    \includegraphics[width=\linewidth]{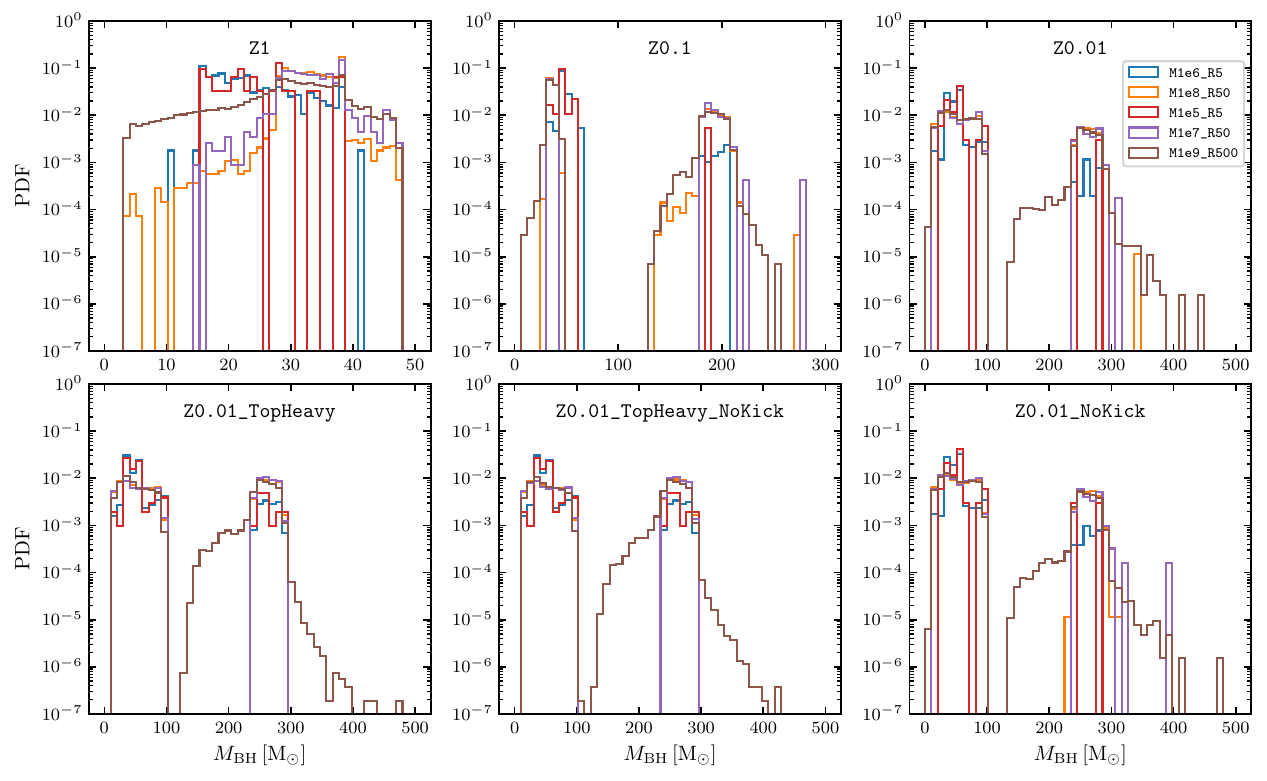}
    \caption{{Mass spectrum of remnant BHs at the end of the simulation for different simulations. The panels represent simulation runs with different metallicity ($0.01\,Z_\odot$--$Z_\odot$), IMF shape, and natal kicks (as labeled therein). In each panel are the five simulation runs with their GMC initial conditions listed in Table~\ref{tab:deluxesplit}. We caution that only the BH remnants of the resolved stars in the simulations are shown.
    }
    }
    \label{fig:hist_bh_mass}
\end{figure*}

\begin{figure*}
    \centering
    \includegraphics[width=\linewidth]{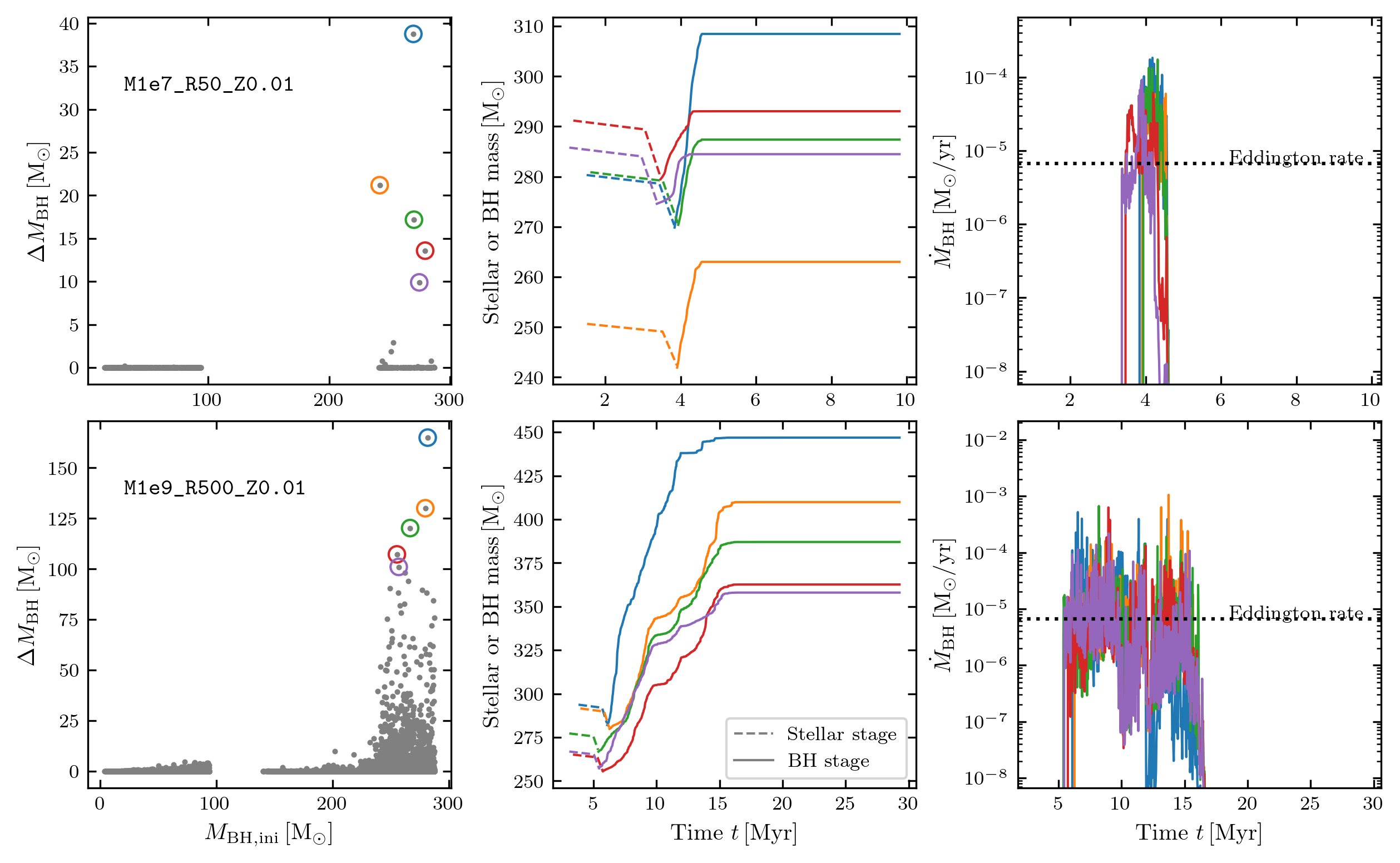}
    \caption{Mass accretion history of selected remnant BHs. Here we present two simulation runs, \path{M1e7_R50_Z0.01} (\emph{top}) and \path{M1e9_R500_Z0.01} (\emph{bottom}). \emph{Left.}--Remnant BH mass at formation versus the accreted mass, where 5 BHs with the most accretion are emphasized with colored circles. \emph{Middle.}--The mass evolution of the selected stars and their remnants (matching the colors to the left panels). \emph{Right.}--The accretion rate of selected BHs, compared with the Eddington accretion rate of a $300\,M_\odot$ BH (\emph{black dotted}).
    }
    \label{fig:bh_mass_growth}
\end{figure*}

\begin{figure}
    \centering
    \includegraphics[width=\linewidth]{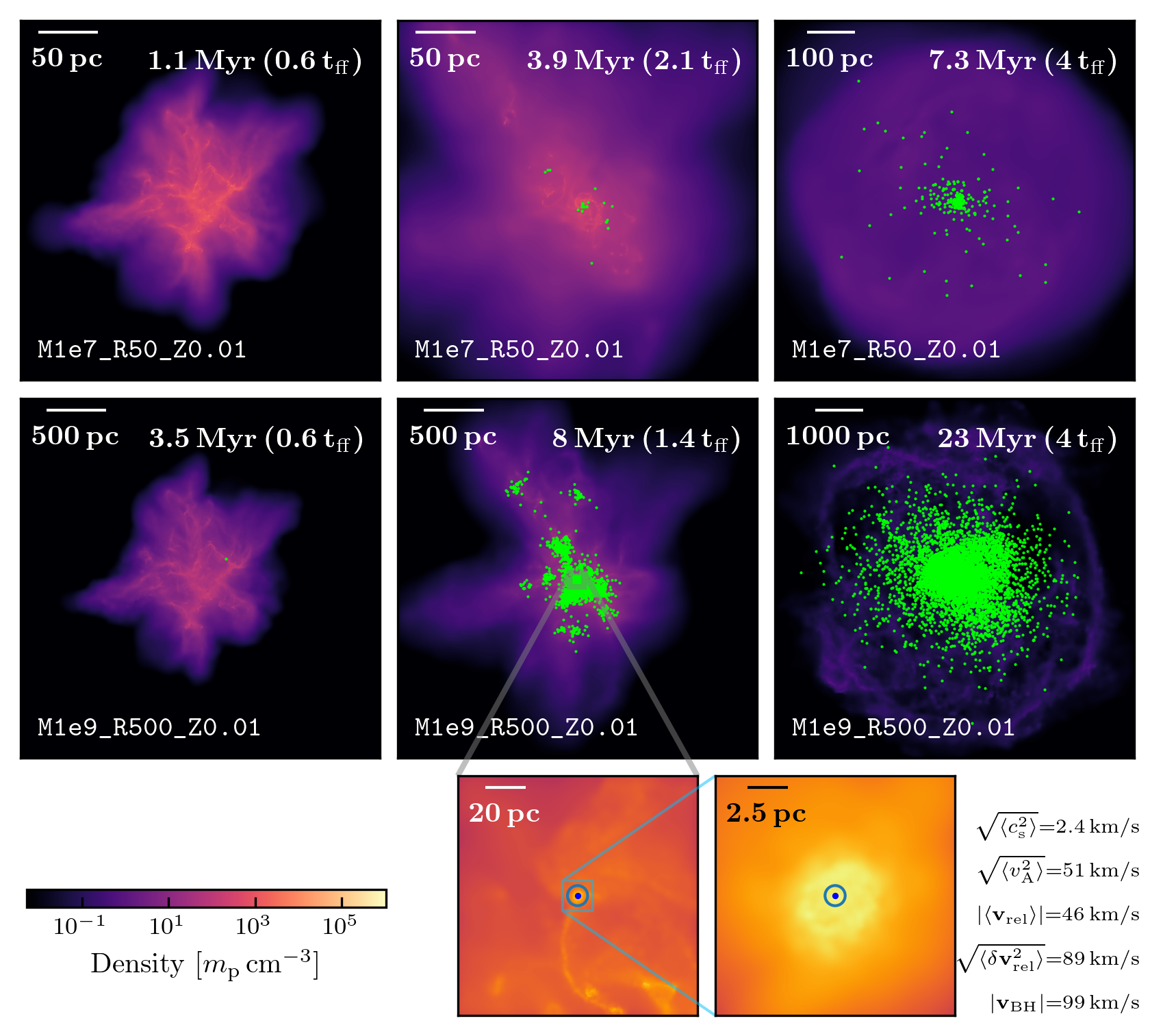}
    \caption{{Visualization of gas morphology and positions of BHs (green or dark dots) in simulations. \emph{Top row.}-- Simulation \path{M1e7_R50_Z0.01} at different stages (left to right: early collapse of the GMC; remnant BH appearance; disruption of the gas reservoir). \emph{Bottom row and inset panels.}-- Similar but for \path{M1e9_R500_Z0.01}. The inset panels focuses on a BH undergoing rapid accretion, which is embedded in a turbulent, magnetized, dense gaseous core (see quantities listed to the right). }
    }
    \label{fig:visualization}
\end{figure}

\subsection{Simulations}

We construct the initial conditions (ICs) for our giant molecular cloud (GMC) simulations following the methodology described in \citet{GrudicGuszejnovHopkins_2018MNRAS.481..688G} and \citet{ShiKremerGrudic_2023MNRAS.518.3606S,ShiKremerHopkins_2024A&A...691A..24S}. Each simulation begins with a spherical cloud IC \emph{without} any pre-existing BHs. The ICs are non-rotating and uniform in density, but are seeded with a Gaussian turbulent velocity field in which $50\%$ of the modes are solenoidal. The initial magnetic field is uniform, but becomes turbulent as the simulation evolves due to gas motions. The initial kinetic and magnetic energies are set to $100\%$ and $1\%$ of the initial gravitational binding energy, respectively. Each cloud is initially $10^4\,\rm K$, then it cools radiatively, undergoes gravitational collapse, and forms stars. Each simulation is evolved for a duration of $\max(10\,{\rm Myr}, 5\,t_{\rm ff})$.

\begin{deluxetable}{cccccc}
\tablewidth{0pt} 
\tablecaption{GMC initial conditions of the simulations suite. \label{tab:deluxesplit}}
\tablehead{
\colhead{$Z_{\rm ini}/Z_\odot$} & \colhead{$\bar \Sigma_0$}& \colhead{$R_{\rm cl}$} & \colhead{$M_{\rm cl}$} & $t_{\rm ff}$ & $m_{\rm gas}$   \\
 & \colhead{($M_\odot\, {\rm pc}^{-2}$)} & \colhead{($\rm
pc$)}& \colhead{($M_\odot$)} & \colhead{(Myr)} & \colhead{($M_\odot$)} 
} 
\colnumbers
\startdata 
\multirow{5}{*}{1, 0.1, 0.01}
& \multirow{2}{*}{$1.3\times 10^4$}& 5 & $10^6$ & 0.185 & 3.8  \\ 
& & 50 &    $10^8$& 0.585  & 380 \\ 
\cline{2-6}
& \multirow{3}{*}{$1.3\times 10^3$}& 5 & $10^5$  & 0.585 & 0.38  \\ 
& & 50 &  $10^7$& 1.85  & 38
 \\ 
& & 500 & $10^{9}$& 5.85  & 3800
\enddata
\tablecomments{Columns of the table: (1) initial metallicity; (2) initial mean surface density; (3) initial radius; (4) initial mass; (5) initial free-fall time; (6) mass resolution of gas. }
\end{deluxetable}

We perform a suite of simulations designed to isolate the impact of several key physical parameters.

\begin{itemize}
\item \emph{GMC properties.} The initial surface density of a GMC is the primary factor controlling the star formation efficiency and BH accretion \citep[e.g.,][]{GrudicGuszejnovHopkins_2018MNRAS.481..688G,ShiKremerGrudic_2023MNRAS.518.3606S}. Accordingly, we fix the initial surface density to $\Sigma \sim 10^4\,M_\odot\,{\rm pc}^{-2}$ and $\sim 10^3\,M_\odot\,{\rm pc}^{-2}$, while varying the cloud radius over ${5,50,500}\,\rm pc$. See Table~\ref{tab:deluxesplit}.

\item \emph{Initial metallicity.} Metallicity influences both star formation and feedback through its effects on gas cooling and the coupling between radiation and gas. In addition, stellar remnants are metallicity-dependent \citep{CostaShepherdBressan_2025A&A...694A.193C}, which directly affects the masses of compact objects.

\item \emph{IMF.} By default, we adopt a Kroupa IMF, with $\dd n/\dd m \propto m^{-2.3}$ for $m>0.5\,M_\odot$. Motivated by star formation simulations suggesting a top-heavy IMF at low metallicity (\citealt{ChonOmukaiSchneider_2021MNRAS.508.4175C}), we also consider a variant of $\dd n/\dd m \propto m^{-1.3}$. 

\item \emph{Natal kicks.} Natal kicks are included by default in our simulations. At $Z=0.01\,Z_\odot$, we additionally perform runs without natal kicks for comparison.
\end{itemize}

Taking these factors into account, our simulation suite comprises five distinct ICs with different masses and radii (Table~\ref{tab:deluxesplit}), each further varied by metallicity, IMF, and natal-kick prescriptions. For notational convenience, we denote an IC by \texttt{M\%g\_R\%g\_Z\%g}, indicating its mass (\texttt{M}, in $M_\odot$), radius (\texttt{R}, in pc), and initial metallicity (\texttt{Z}, in $Z_\odot$). Additional suffixes, such as \texttt{TopHeavy} or \texttt{NoKick}, denote simulations adopting a top-heavy IMF or excluding natal kicks, respectively.

\section{Results}
\label{sec:results}

\subsection{Seeding of black holes}

Figure~\ref{fig:sfr-bhfr} shows the star formation rate (SFR) and BH formation rate (BHFR, defined as the $\dd M_{\rm BH,tot}/\dd t$) in GMCs with different masses and radii. For each IC, variations in metallicity, IMF, and natal kicks are displayed in the same panel. 

Comparing the SFR in absolute time (in Myr) and free-fall times ($t_{\rm ff}$), there is a rough consistency among simulations from different ICs that SFR peaks at $\sim t_{\rm ff} - 2 t_{\rm ff}$ after the start of GMC's gravitational collapse. Moreover, the SFR of simulations from different ICs drops below $10^{-4}\,M_\odot\,{\rm yr}^{-1}$ after $\sim 5\,t_{\rm ff}$, which marks the end of star formation in this complex. This is in agreement with other simulations of star formation in single GMCs \citep{GrudicHopkinsFaucher-Giguere_2018MNRAS.475.3511G}, i.e., the peak SFR is $\sim \epsilon_{\rm ff} M_{\rm gas}/t_{\rm ff}$, where $\epsilon_{\rm ff} \equiv M_{\star}/M_{\rm gas}$ is the star formation efficiency. However, since BHs only form after the death of massive stars, the BHFR keeps a similar overall shape as SFR but is delayed by $\gtrsim 3\,\rm Myr$. 

Figure~\ref{fig:sfr-bhfr} also reflects the impact of different physics on star formation and BH formation. By including a top-heavy IMF, more massive stars are generated in the simulation, boosting both the radiative and mechanical feedback, since the stellar population will have higher luminosities (and light-to-mass ratio) and higher mass loss rate \citep{Vink_2023A&A...679L...9V} than those with the canonical IMF. In Figure~\ref{fig:sfr-bhfr}, the top-heavy IMF suppresses the overall SFR, but the corresponding BHFR is much higher than that with the canonical IMF, implying a much larger BH population due to the overabundance of massive stars. Natal kicks of BHs do not noticeably change the SFR or BHFR in the simulations.

Figure~\ref{fig:density-profile} shows the spherical density profile of the major star cluster at the end of the simulations, with the center selected as the minimum of the gravitational potential \citep{GrudicGuszejnovHopkins_2018MNRAS.481..688G}. We note that not all simulations generate a regular-shaped spherical globular cluster; we find messy ``open clusters'' in a few cases. Some clusters do not have a well-defined center, and the mass density profile is noisy. Comparing two groups of simulations with different initial surface densities, we find that the group with higher surface density $\sim 10^4\,M_\odot\,{\rm pc}^{-2}$ typically forms a massive central globular cluster at the center (panels a, b), while it is not common for the group with $\bar\Sigma_0\sim 10^3\,M_\odot\,{\rm pc}^{-2}$, especially for low-mass clouds (panels c, d, e). The central densities of stars and BHs are both high for the group with higher surface density, reaching $\sim 10^{6-7}\,M_\odot\, {\rm pc}^{-3}$ for stars and $\sim 10^5\,M_\odot\, {\rm pc}^{-3}$ for BHs. The outer density profile of all clusters is well approximated by $\rho(r)\propto r^{-7/2}$.

{Interestingly, the inclusion of a top-heavy IMF significantly changes the density profile, making it ``puffier'': both the central density and density normalization are lower than the runs with a standard IMF, or even failing to form a globular cluster for \path{M1e5_R5} clouds (red and purple lines). }

We find that natal kicks of BHs can sometimes impact the density profile of the final cluster in low-mass star-forming clouds. For example, in runs with \path{M1e5_R5}, there is a regular-shaped final globular cluster forming at $Z=0.01\,Z_\odot$ only when the natal kick is excluded (Figure~\ref{fig:density-profile}). This is probably due to the high mass fraction of BHs in the stars.

Figure~\ref{fig:hist_bh_mass} shows the mass spectrum of BHs at the end of the simulation. The mass spectrums show metallicity dependence \citep{CostaShepherdBressan_2025A&A...694A.193C}: at $Z_\odot$, there is only a single group of BHs that is $\lesssim 50\,M_\odot$; at lower metallicities ($0.1\,Z_\odot$, $0.01\,Z_\odot$), there are two groups separated by the PISN mass gap.

The mass spectrum also showcases whether there is accretion onto the remnant BHs. For $Z=Z_\odot$ and $Z=0.1\,Z_\odot$, the maximum possible BH mass is well below $300\,M_\odot$, the preset maximum stellar mass, suggesting that the accretion (if any) is inefficient. The case is different for $Z=0.01\,Z_\odot$, where some of the runs (particularly \path{M1e9_R500}) have BH masses $>300\,M_\odot$, suggesting that there is significant accretion. Still, these BHs are $\lesssim 500\,M_\odot$, which are not ``heavy seeds.''

The mass spectrum also hints at the effect of other physics. A top-heavy IMF generates more massive remnants, but the most massive BH is still $\sim 400-500\,M_\odot$. When natal kicks are not considered, the BHs gain slightly more accretion: e.g., with natal kicks, the heaviest BH in the \path{M1e9_R500_Z0.01} run is $\sim 450\,M_\odot$; while it is $\sim 475\,M_\odot$ in the run without natal kicks.

\subsection{Accretion of seed black holes}

From our simulations that explicitly track the lifetimes of VMSs and their remnant BHs, we find that a small subset of remnant BHs grow to $\sim 400$--$500\,M_\odot$. This occurs only at low metallicity ($Z=0.01\,Z_\odot$) in massive and extended clouds. Representative examples include \path{M1e9_R500_Z0.01} and \path{M1e7_R50_Z0.01}. Below, we examine the accretion histories of BHs in these clouds.

Figure~\ref{fig:bh_mass_growth} shows the mass evolution of the five BH seeds with the largest growth in each simulation. These BHs all lie above the pair-instability mass gap, as remnants of low-metallicity ($0.01\,Z_\odot$) and massive ($\gtrsim 200\,M_\odot$) progenitor stars. After a stellar evolution phase (\emph{dashed}) lasting $\sim 3\,\rm Myr$, the stars collapse into BHs (\emph{solid}) and undergo a short episode of super-Eddington accretion. In the \path{M1e7_R50_Z0.01} run, BHs accrete at rates of $\sim 10^{-4}\,M_\odot\,{\rm yr}^{-1}$, corresponding to $\sim 10\,\dot M_{\rm Edd}$, for $\sim 1\,\rm Myr$, producing $\sim 300\,M_\odot$ black holes. In contrast, in \path{M1e9_R500_Z0.01}, comparable accretion rates persist for $\sim 10\,\rm Myr$, allowing BHs to grow to $\sim 450\,M_\odot$. In both cases, the termination of accretion coincides with the decline of star formation activity (Figure~\ref{fig:sfr-bhfr}), which is driven by the depletion of the dense gas reservoir.

A common property of these rapidly growing BHs (Figure~\ref{fig:bh_mass_growth}) is that they form via direct collapses, with minimal energy injection at the collapse stage (while retaining nearly all of the pre-SN stellar mass), which prevents early evacuation of the surrounding dense gas.

Figure~\ref{fig:visualization} compares simulations with and without significant BH accretion. In the \path{M1e7_R50_Z0.01} run, the first remnant BHs form at $\sim 2\,t_{\rm ff}$, after the peak of star formation has passed (cf. Figure~\ref{fig:sfr-bhfr}). Subsequent SN explosions expel gas to scales $\gtrsim 100\,\rm pc$, leaving little opportunity for sustained BH accretion. In contrast, in the \path{M1e9_R500_Z0.01} run, dense gas remains abundant and star formation is still ongoing at $\sim 1\,t_{\rm ff}$, while a substantial population of remnant BHs has already formed. Insets focusing on the most rapidly accreting BHs show that they are embedded within dense cores with $n \sim 10^{5}$--$10^{6}\,\rm cm^{-3}$.

This accretion environment closely resembles that identified in \citet{ShiKremerGrudic_2023MNRAS.518.3606S,ShiKremerHopkins_2024A&A...691A..24S}. For gas cells in these dense cores, the BH--gas relative velocity $\mathbf{v}_{\rm rel} = \mathbf{v}_{\rm BH} - \mathbf{v}_{\rm gas}$ satisfies $c_{\rm s} \lesssim |\langle \mathbf{v}_{\rm rel} \rangle| \ll |\mathbf{v}_{\rm BH}| \sim v_{\rm circ}$, where $v_{\rm circ} = \sqrt{G M_{\rm cl}/R_{\rm cl}}$ is the circular velocity of the GMC. The gas is both magnetized and turbulent, with $v_{\rm A} > c_{\rm s}$ and $|\langle \mathbf{v}_{\rm rel} \rangle| \ll \sqrt{\langle \delta  \mathbf{v}_{\rm rel}^2 \rangle}$.

{By $\sim 4\,t_{\rm ff}$ (see the right panels of Figure~\ref{fig:visualization}), a dense star cluster forms in both \path{M1e7_R50_Z0.01} and \path{M1e9_R500_Z0.01}. As stars, SNe, and BHs inject substantial radiative and mechanical energy into the surrounding gas, and the remaining gas reservoir is expelled, abruptly halting BH accretion (as in Figure~\ref{fig:bh_mass_growth}).
}

\section{Discussion}
\label{sec:discussion}

\begin{figure}
    \centering
    \includegraphics[width=\linewidth]{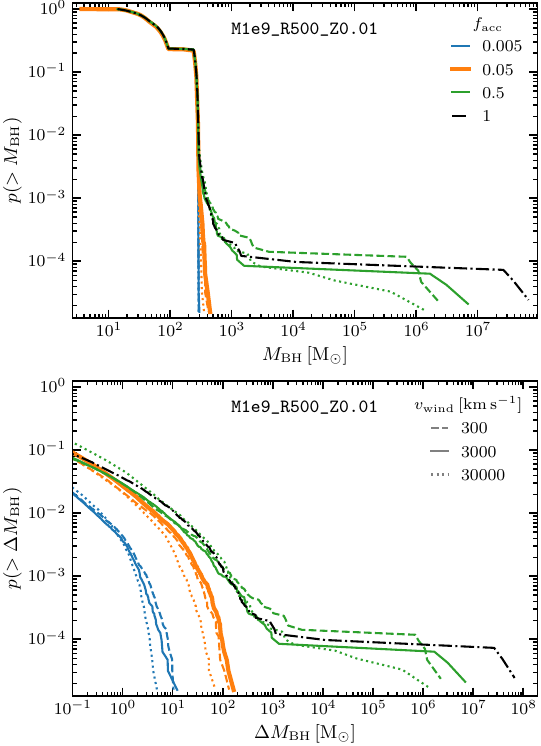}
    \caption{{Tests with BH mechanical feedback models based on the cloud \path{M1e9_R500_Z0.01}. We plot the cumulative distribution of the final BH mass $M_{\rm BH}$ (\emph{top}) and the accreted mass $\Delta M_{\rm BH}$ of each BH (\emph{bottom}). For each test, we keep all fiducial setups except $f_{\rm acc}$ (0.005, 0.05, 0.5; {colored lines}) and $v_{\rm wind}$ (300, 3000, $3000\,\rm km\,{s}^{-1}$). Additionally, we test with $f_{\rm acc}=1$ and all BH feedback disabled (\emph{black dot dashed}).}
    }
    \label{fig:feedback_test}
\end{figure}

\begin{figure*}
    \centering
    \includegraphics[width=\linewidth]{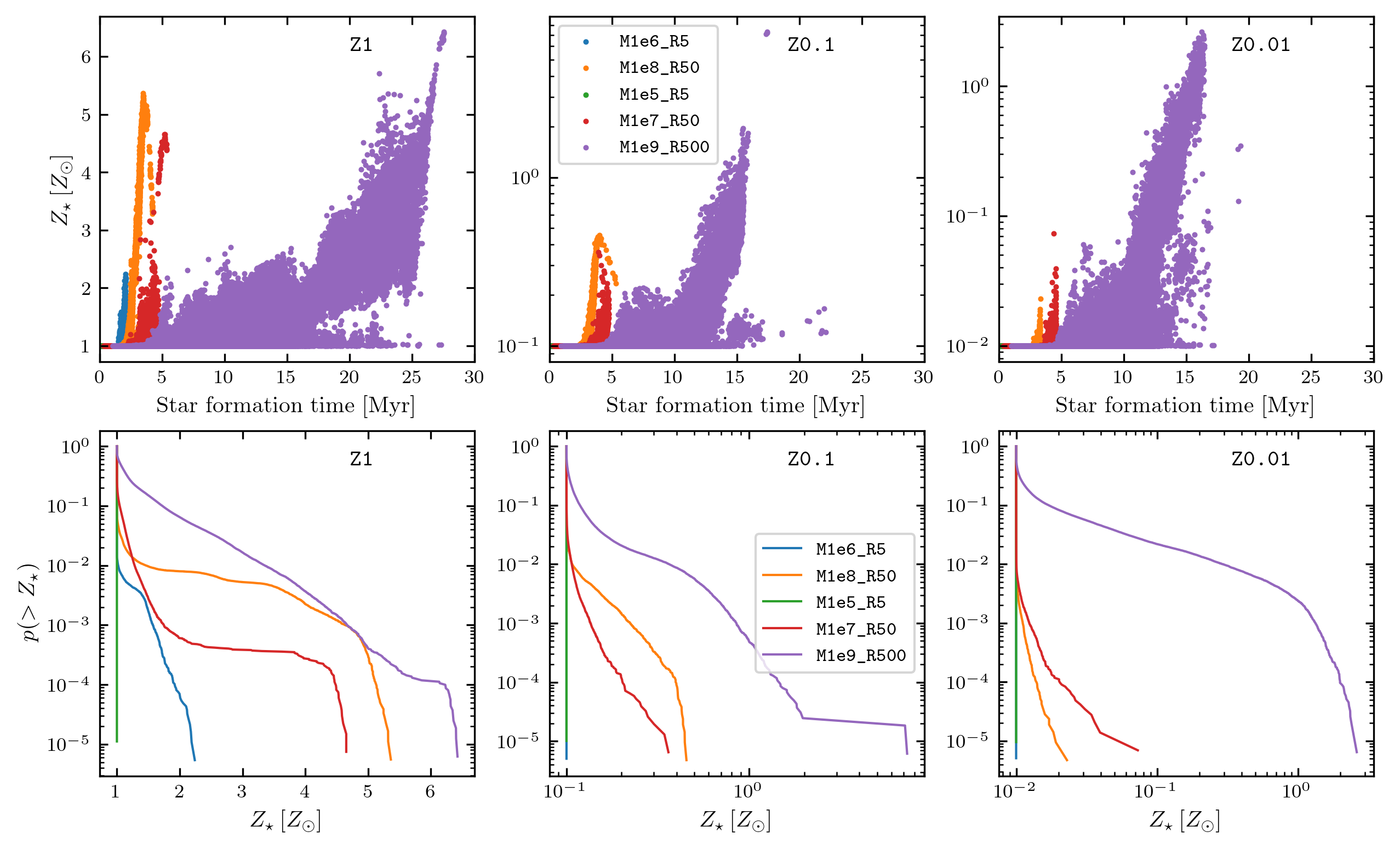}
    \caption{{The metallicity of stars at the end of each simulation. \emph{Top.}--The {scatter} plot of the star's metallicity at formation ($Z_\star$) versus their formation time, for clouds of $Z_\odot$ (\path{Z1}; \emph{left}), $0.1\,Z_\odot$ (\path{Z0.1}; \emph{middle}), and $0.01\,Z_\odot$ (\path{Z0.01}; \emph{right}). Dots with different colors represent different GMC initial conditions listed in Table~\ref{tab:deluxesplit}. \emph{Bottom.}--Cumulative distribution of stellar metallicity $Z_\star$ at the end of the simulation.} }
    \label{fig:sftime_metallicity}
\end{figure*}

\subsection{The sub-grid BH feedback model}
\label{sec:discussion:feedback_test}

As introduced in \S~\ref{sec:method}, our BH feedback model includes both radiative and mechanical components. However, this sub-grid prescription involves uncertainties, particularly in the fraction $f_{\rm acc}$ of the Bondi accretion rate $\dot M_{\rm Bondi}$ that reaches the event horizon, and in the BH outflow velocity $v_{\rm wind}$. Both parameters are poorly constrained and lack direct calibration. To assess their impact on BH growth, we perform a set of additional simulations using the \path{M1e9_R500_Z0.01} cloud.

Figure~\ref{fig:feedback_test} summarizes the results by showing the cumulative distributions of the final BH mass $M_{\rm BH}$ and the total accreted mass $\Delta M_{\rm BH}$. For runs with $f_{\rm acc}=0.005$, varying $v_{\rm wind}$ does not produce any massive BHs; as shown in the bottom panel, the accreted mass is at most $\sim 10\,M_\odot$. When $f_{\rm acc}=0.05$, corresponding to our fiducial value, the most massive BH reaches $\lesssim 500\,M_\odot$, and the accreted mass is at most $\sim 100\,M_\odot$, depending on the assumed value of $v_{\rm wind}$. In this regime, higher wind velocities (e.g., $3000\,\rm km\,s^{-1}$) further suppress BH accretion. Comparing the cumulative distributions of $\Delta M_{\rm BH}$ for runs with $f_{\rm acc}=0.005$ and $f_{\rm acc}=0.05$, we find a factor of $\sim 10$ difference, as expected from the linear scaling implied by the sub-grid accretion model.

The most striking behavior occurs in the case with $f_{\rm acc}=0.5$, where a large fraction of the Bondi rate is assumed to reach the BH. In these runs, accretion produces massive BHs with $M_{\rm BH} \sim 10^6$--$10^7\,M_\odot$. The cumulative mass distribution exhibits a plateau between $\sim 10^3\,M_\odot$ and $\sim 10^6\,M_\odot$, indicating that only $\sim 3$ BHs reach masses of $\sim 10^6\,M_\odot$, while the majority of the BH population remains below $\sim 10^3\,M_\odot$. This behavior can be understood in terms of Bondi accretion, which scales as $\dot M_{\rm Bondi} \propto M_{\rm BH}^2$, such that more massive BHs experience rapidly increasing accretion rates and can undergo runaway growth. These results suggest the presence of a characteristic mass threshold of $\sim 10^3\,M_\odot$ for runaway accretion: in runs with $f_{\rm acc}=0.005$ and $f_{\rm acc}=0.05$, all BHs remain below this threshold, whereas for $f_{\rm acc}=0.5$, a small number of BHs are able to cross it and subsequently experience runaway growth.

Finally, we consider an extreme simulation in which all BH feedback is disabled, corresponding to $f_{\rm acc}=1$. This idealized case represents the maximum possible accretion scenario. In this run, the most massive BH reaches $\sim 10^8\,M_\odot$, although only a very small fraction of the BH population ($\sim 10^{-4}$) undergoes runaway accretion. {Comparing this no-feedback $f_{\rm acc}=1$ simulation with the simulations of $f_{\rm acc}=0.5$, we find that the accreted mass of most BHs is similar, while the mass of the most massive few BHs is suppressed in the $f_{\rm acc}=0.5$ runs due to BH feedback. This may suggest that for most BHs, the BH feedback is insufficient to disrupt the dense gaseous cores around them, while only the most massive BHs are strongly regulated by their feedback. 
}

In summary, we find that $f_{\rm acc}$ is a critical parameter governing remnant BH growth from our tests with the \path{M1e9_R500_Z0.01} cloud. For $f_{\rm acc}\lesssim 0.5$, the accreted mass $\Delta M_{\rm BH}$ scales approximately linearly with $f_{\rm acc}$, and no heavy seeds form. Conversely, for $f_{\rm acc}\gtrsim 0.5$, a small number of ``lucky'' stellar-mass BHs can grow rapidly to masses above $\sim 10^3\,M_\odot$, enabling them to attract additional gas and enter a phase of runaway accretion towards $\sim 10^6\,M_\odot$ SMBHs. From Figure~\ref{fig:feedback_test}, given that radiative feedback coexists, the BH wind's terminal velocity also affects BH accretion slightly: accretion is more suppressed for $v_{\rm wind}=30000\,\rm km\,s^{-1}$ than $v_{\rm wind}=3000\,\rm km\,s^{-1}$ and $v_{\rm wind}=300\,\rm km\,s^{-1}$. An exception is the $\sim 3$ most massive BHs in each of the unrealistic $f_{\rm acc}=0.5$ runs, which may reflect small-number statistics; the remaining BH population, however, follows the overall trend.

{As a caveat, we note that the runaway accretion in our simulations could also be related to the mass dependence of the BH accretion rate estimator, $\dot M_{\rm BH} \propto M_{\rm BH}^\beta$. For the BHL formula used here, the mass dependence is steep ($\beta=2$). However, it remains possible that for models with shallower mass dependence (smaller $\beta$), like the gravitational-torque-driven accretion \citep{HopkinsQuataert_2011MNRAS.415.1027H,Angles-AlcazarDaveFaucher-Giguere_2017MNRAS.464.2840A,Angles-AlcazarQuataertHopkins_2021ApJ...917...53A} and free-fall-intuited sink prescriptions \citep{WeinbergerBhowmickBlecha_2025A&A...700A..52W}, runaway accretion of massive BHs may be suppressed.
}

\subsection{Multiple-generation star formation}

We find signatures of multiple-generation star formation in these simulations, i.e., young stellar populations are born from the gas that is chemically enriched by the previous generation of stars. One example is provided by the mass spectrum of stellar-mass BHs shown in Figure~\ref{fig:hist_bh_mass}. At $Z=0.01,Z_\odot$, the \path{M1e9_R500} cloud exhibits a population of BHs in the mass range of $\sim 120$--$250\,M_\odot$, which is absent in all other clouds. This sub-population could originate from stars formed at metallicities $Z>0.01\,Z_\odot$, enabled by the presence of multiple generations of star formation.

Figure~\ref{fig:sftime_metallicity} shows $Z_\star$, the metallicity of stars at their formation, in different simulations. The top row presents the scatter plot between the star formation time and $Z_\star$. For each cloud, $Z_\star$ is equal to the initial metallicity $Z_{\rm ini}$ at early times; subsequently, the metallicity of newly formed stars increases with time. This trend is particularly pronounced in the \path{M1e9_R500} cloud, where a large number of stars form with $Z \gg Z_{\rm ini}$. For $Z_{\rm ini}=Z_\odot$, high-metallicity stars can reach $Z_\star \sim 6\,Z_\odot$. Even in low-metallicity clouds with $Z_{\rm ini}=0.01\,Z_\odot$, stars may form at high metallicities like $Z_\star \sim 2\, Z_\odot$, suggesting significant metal enrichment from evolved stars.

The bottom row shows the cumulative distribution of $Z_\star$ at the end of the simulation. Comparing different clouds, we find that \path{M1e9_R500} exhibits the highest level of multiple-generation star formation, conversely all stars in \path{M1e5_R5} have the same metallicity. For example, in the run with $Z_{\rm ini}=0.01\,Z_\odot$, $\sim 10\%$ of the stars are formed at $Z > 2\,Z_{\rm ini}$. This behavior can be understood in the context of the star formation history shown in Figure~\ref{fig:sfr-bhfr}: when the star formation rate peaks at $\sim 1\,t_{\rm ff} \sim 7\,\rm Myr$, the cluster retains a high fraction of dense gas while first-generation stars are dying and injecting chemical feedback. Other clouds, however, are unable to retain dense gas during the epoch of chemical feedback, resulting in a lower level of multiple-generation star formation.

The existence of high-metallicity, second-generation stars implies efficient retention of (chemically enriched) dense gas, which shares the same logic as that of significant BH remnants. Consequently, there could be an intrinsic link between remnant BH accretion and multiple-generation star formation. From our simulations, clouds (e.g., \path{M1e9_R500_Z0.01}) with a higher level of multiple-generation star formation also produce heavier BHs. 

These results also suggest the possibility of VMSs in explaining the multiple population problem of massive globular clusters \citep{BastianLardo_2018ARA&A..56...83B,MiloneMarino_2022Univ....8..359M}, which features the star-by-star spread in $\alpha$-element abundance. More careful star formation simulations with VMSs, particularly in the cosmological context, could help assess this argument.

\subsection{Can light seeds become SMBHs?}

Observations have identified molecular clouds with high surface densities at high redshift \citep{FujimotoOuchiKohno_2025NatAs...9.1553F}, and $\sim 10^7\,M_\odot$ massive globular cluster at cosmic noon \citep{PascaleDaiMcKee_2023ApJ...957...77P}. Star formation in such environments inevitably produces numerous massive remnant BHs. Although our simulations show that early, in-situ growth of light seeds is inefficient due to gas depletion by stellar feedback (even in dense environments), the possibility of these BHs being seeds of SMBHs during their sequential evolution should not be excluded.

One possibility is renewed gas accretion driven by subsequent large-scale inflows. In this case, the remnant BHs effectively become pre-existing seeds, similar to the scenario explored in \citet{ShiKremerHopkins_2024ApJ...969L..31S}. If such seeds reside in the deep potential well of a dense globular cluster, potentially augmented by a dark matter halo, infalling gas can provide a significant mass supply for BH accretion, although this feeding cycle can be interrupted by another wave of starburst and stellar feedback \citep{SunseriAndalmanTeyssier_2025arXiv251019822S}. Similar scenarios are addressed by several recent studies \citep{PartmannNaabLahen_2025MNRAS.537..956P,ReinosoLatifSchleicher_2025A&A...700A..66R,MukherjeeZhouChen_2025ApJ...981..203M}.

Another pathway arises from the extremely high stellar and BH densities reached in some clusters (Figure~\ref{fig:density-profile}), which enhance two-body interactions such as stellar collisions, tidal disruption events, and BH-BH mergers. Repeated tidal disruptions followed by accretion can efficiently increase IMBH masses, particularly in clusters with short core-collapse times \citep{RantalaNaabLahen_2024MNRAS.531.3770R}. Runaway stellar mergers may also lead to supermassive stars and IMBHs \citep{ReinosoLatifSchleicher_2025A&A...700A..66R}. {Finally, through the cosmic hierarchical assembly histroy of galaxies, repeated mergers associated with stellar-mass BH can further boost the masses of central IMBHs and SMBHs \citep{KritosBeckmannSilk_2025ApJ...991...58K}. IMBHs may also undergo repeated mergers along with galaxy mergers to form SMBHs \citep{BhowmickBlechaTorrey_2025MNRAS.538..518B,BhowmickBlechaTorrey_2026ApJ...997..187B}, and {subsequently} build up their masses through gas accretion \citep[e.g.,][]{ZhouDiMatteoBird_2026ApJ...999...41Z}.
}

\section{Conclusions}
\label{sec:summary}

In this study, we adopt a ``FIRE+VMS'' framework (Figure~\ref{fig:diagram}) to simulate the in-situ formation of stellar-mass BHs (``light seeds''). Through star formation in massive GMCs, massive and compact clusters, which are analogous to high-redshift nuclear star clusters, are formed. Light seeds are produced directly through stellar evolution (instead of being pre-existing), and their feedback-regulated accretion of the leftover, in-situ gas is carefully modeled to inspect if any IMBHs can form. 

Our simulations span a broad parameter space of GMC's initial properties (Table~\ref{tab:deluxesplit}) and consider complexities like natal kicks and BH feedback. Below are our general conclusions regarding BH seeding and growth.
\begin{enumerate}
    \item With the fiducial feedback-regulated BH accretion model included, no ``heavy seeds'' form within the $\sim 10\,\rm Myr$ lifetime of GMCs, due to the fact that dense gas is efficiently removed by continued stellar feedback including radiation, winds, and SN explosions.
    
    \item The star formation rate peaks at $\sim (1$--$2)\,t_{\rm ff}$, while remnant BH formation is delayed by $\sim 3\,\rm Myr$. Notable BH accretion only occurs for clouds with $t_{\rm ff}\gtrsim 3\,\rm Myr$ and high surface densities $\gtrsim 10^3\,M_\odot\,{\rm pc}^{-2}$, which still retains gas at BH formation. These clouds are among most massive and extended ones ($M_{\rm cl}=10^9\,M_\odot$, $R_{\rm cl}=500\,\rm pc$).
    
    \item In massive and low-metallicity ($Z_{\rm ini}=0.01\,Z_\odot$) clouds, a small subset of BHs grow from $\lesssim 300\,M_\odot$ to $\sim 400$--$500\,M_\odot$, all of which are massive DCBHs above the PISN mass gap. The resulting accretion can be super-Eddington within cold, magnetized, dense cores \citep[similar to][]{ShiKremerGrudic_2023MNRAS.518.3606S}.
    
    \item A top-heavy IMF increases the number of massive stars and remnant BHs, but enhanced stellar feedback suppresses the overall star formation efficiency. Clusters formed under a top-heavy IMF are correspondingly more extended (``puffier''). Despite the increased number of light seeds, none of them become heavy seeds.
    
    \item Natal kicks of $0.1$--$100\,\rm km\,s^{-1}$ leaves insignificant impact on remnant BH accretion or the global evolution of star-forming complexes, except in low-mass clusters with shallow potential wells.
    
    \item The fraction of Bondi inflow reaching the BH, $f_{\rm acc}$, is a key sub-grid parameter that determines remnant BH accretion. For $f_{\rm acc}\lesssim 0.05$, no massive BHs form even for low outflow velocities. In contrast, for $f_{\rm acc}\gtrsim 0.5$, runaway accretion can occur, allowing a few BHs to reach $\gtrsim 10^6\,M_\odot$.

    \item The star-forming complexes with notable remnant BH accretion also undergo multiple-generation star formation, from which stars of a wide range of metallicities are formed, suggesting a potential, intrinsic connection between these two processes.
\end{enumerate}

{Overall, our fiducial simulations suggest limited accretion of in-situ gas onto newly formed stellar-mass BHs. The fact that the growth of these BHs is limited by stellar feedback is in alignment with previous studies \citep[e.g.,][]{DuboisVolonteriSilk_2015MNRAS.452.1502D,HabouzitVolonteriDubois_2017MNRAS.468.3935H,MehtaReganProle_2024OJAp....7E.107M,ShinSmithSijacki_2026MNRAS.548ag580S}.} Nevertheless, the possibility that such BHs may eventually grow into SMBHs cannot be excluded, given the limitations of this study. For example, the simulations are not embedded in a cosmological context (they follow only $\sim 10\,\rm Myr$) and therefore neglect gas inflows from larger scales and other environmental effects. {Recent cosmological simulations by \citet{MehtaReganProle_2026NatAs.tmp...21M} do find that $\sim 10^4\,M_\odot$ IMBHs may grow from light seeds. However, their simulations neglected mechanical feedback from the BHs and the mass outflow, both of which may suppress the actual BH growth rate.
}

Our FIRE+VMS framework also assumes a universal IMF, whereas a more realistic IMF may depend on local conditions \citep{GuszejnovHopkinsMa_2017MNRAS.472.2107G}. At low metallicity, gravitational collapse and stellar feedback can further promote accretion onto massive protostars, potentially reshaping the IMF \citep{BegelmanVolonteriRees_2006MNRAS.370..289B,ChonOmukai_2020MNRAS.494.2851C,ChonOmukai_2025MNRAS.539.2561C}. Incorporating these effects will be important for more realistic modeling of massive stars.

A possible follow-up is to perform cosmological zoom-in simulations of light-seed formation and growth. Such simulations would enable a more complete assessment of light-seed accretion over cosmic time and help constrain the origin of seed BHs. We leave this exploration to future work.

\begin{acknowledgments}

We acknowledge the support of the Natural Sciences and Engineering Research Council of Canada (NSERC) under funding reference number 568580.
The computation was performed at the University of Toronto cluster ``Niagara'' and ``Trillium'' supported by SciNet (scinethpc.ca) and the Digital Research Alliance of Canada (alliancecan.ca). This research was supported by the Munich Institute for Astro-, Particle and BioPhysics (MIAPbP) which is funded by the Deutsche Forschungsgemeinschaft (DFG, German Research Foundation) under Germany's Excellence Strategy--EXC-2094--390783311. YS thanks Marta Volonteri for insightful discussions during the visit to MIAPbP.
\end{acknowledgments}

%

\vspace{5mm}


\software{{GIZMO} \citep{Hopkins_2015MNRAS.450...53H}
          }




\bibliography{bib}{}
\bibliographystyle{aasjournal}



\end{document}